\documentclass[conference]{IEEEtran}
\IEEEoverridecommandlockouts
\usepackage{cite}
\usepackage{amsmath,amssymb,amsfonts}
\usepackage{algorithmic}
\usepackage{graphicx}
\usepackage{textcomp}
\usepackage{xcolor}
\usepackage[caption=false,font=footnotesize,subrefformat=parens]{subfig}
\def\BibTeX{{\rm B\kern-.05em{\sc i\kern-.025em b}\kern-.08em
    T\kern-.1667em\lower.7ex\hbox{E}\kern-.125emX}}

\graphicspath{{figures/}}
\begin{document}

\title{Compressed-Sensing based Beam Detection in \linebreak 5G NR Initial Access
}

\author{\IEEEauthorblockN{Junmo Sung and Brian L. Evans}
\IEEEauthorblockA{\textit{Wireless Networking and Communications Group} \\
\textit{The University of Texas at Austin, Austin, TX USA}\\
junmo.sung@utexas.edu, bevans@ece.utexas.edu}
}

\maketitle

\begin{abstract}
To support millimeter wave (mmWave) frequency bands in cellular communications, both the base station and the mobile platform utilize large antenna arrays to steer narrow beams towards each other to compensate the path loss and improve communication performance.  
The time-frequency resource allocated for initial access, however, is limited, which gives rise to need for efficient approaches for beam detection.
For hybrid analog-digital beamforming (HB) architectures, which are used to reduce power consumption, 
we propose a compressed sensing (CS) based approach for 5G initial access beam detection that is for a HB architecture and that is compliant with the 3GPP standard.
The CS-based approach is compared with the exhaustive search in terms of beam detection accuracy and by simulation is shown to outperform. 
Up to 256 antennas are considered, and the importance of a careful codebook design is reaffirmed. 
\end{abstract}


\section{Introduction} 
\label{ch4_sec:introduction}
Directional transmission and reception link has become inevitable in 5G New Radio (NR) due to adoption of millimeter wave (mmWave) frequency bands. As a growing number of antennas are taken into account at base stations (gNB), many issues have been brought up to be addressed. A larger antenna array inherently comes with sharper beams, which in turn more beams are required to sweep and illuminate the entire cell coverage. From the beam management perspective, it implies more frequent beam losses and harder beam recovery. Especially in the initial access process, only a limited opportunity is given to gNB to sweep its possible beams in downlink, e.g., $64$ synchronization signal blocks (SSB) in FR2 \cite{ts38213}. The limitation gives rise to need for an efficient approach to fully exploit the given resources. 

Beam management at mmWave is extensively surveyed from the 3GPP perspectives in \cite{giordani2019tut}. The hybrid beamforming (HB) architecture is shown to be a good trade-off in terms of reactiveness and design complexity. However, the provided results only considers the exhaustive search (ES) for the sake of the standard compliance. Compressive sensing (CS) has been considered as a beam detection method in \cite{choi2015beam,yan2016comp,myers2019loc}, but the literatures lack consideration of the standard and hybrid beamforming architectures. The iterative search is also a good candidate \cite{desia2014init,wei2017exha}, but is not considered in this paper as 
(i) it is shown to exhibit higher misdetection probability in general in \cite{gior2016init}, 
(ii) it requires feedback from users at every iteration to narrow down the search space 
and 
(iii) it is not suitable for initial access due to the user specificness. 

In this paper, 
we propose a CS-based downlink beam detection approach for mmWave hybrid analog and digital beamforming communication systems that is compliant with the 5G NR 3GPP standard. The frequency and time structure of synchronization signal block is specifically taken into account in the design.
The ES is considered as a baseline method which the CS approach is compared against using a beam pair detection probability as a performance metric. 

\section{System and Channel Models} 
\label{ch4_sec:system_and_channel_models}

\begin{figure*}[t!]
  \centering
  \includegraphics[width=17cm]{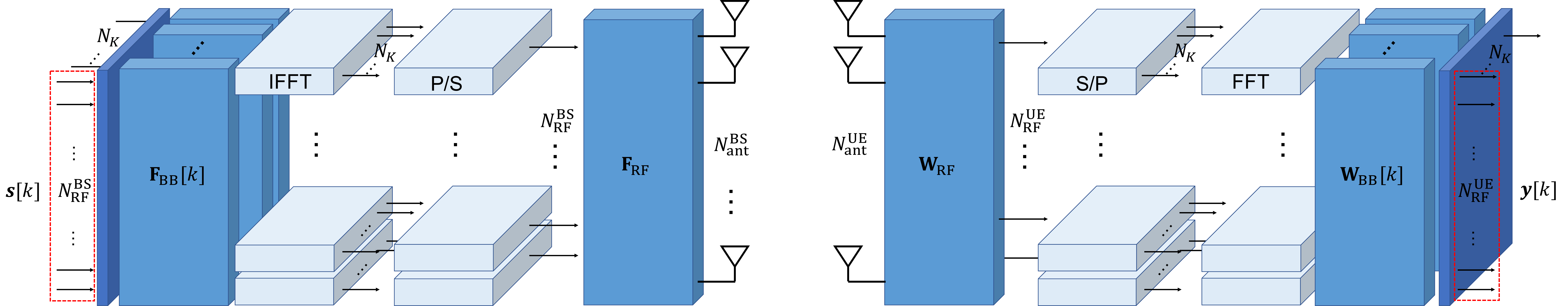}
  \caption{Block diagram of a general OFDM hybrid analog and digital beamforming architecture at both a transmitter and a receiver.}
  \label{ch4_fig:system_diagram}
\end{figure*}
\begin{figure}
  \centering
  \includegraphics[width=7cm]{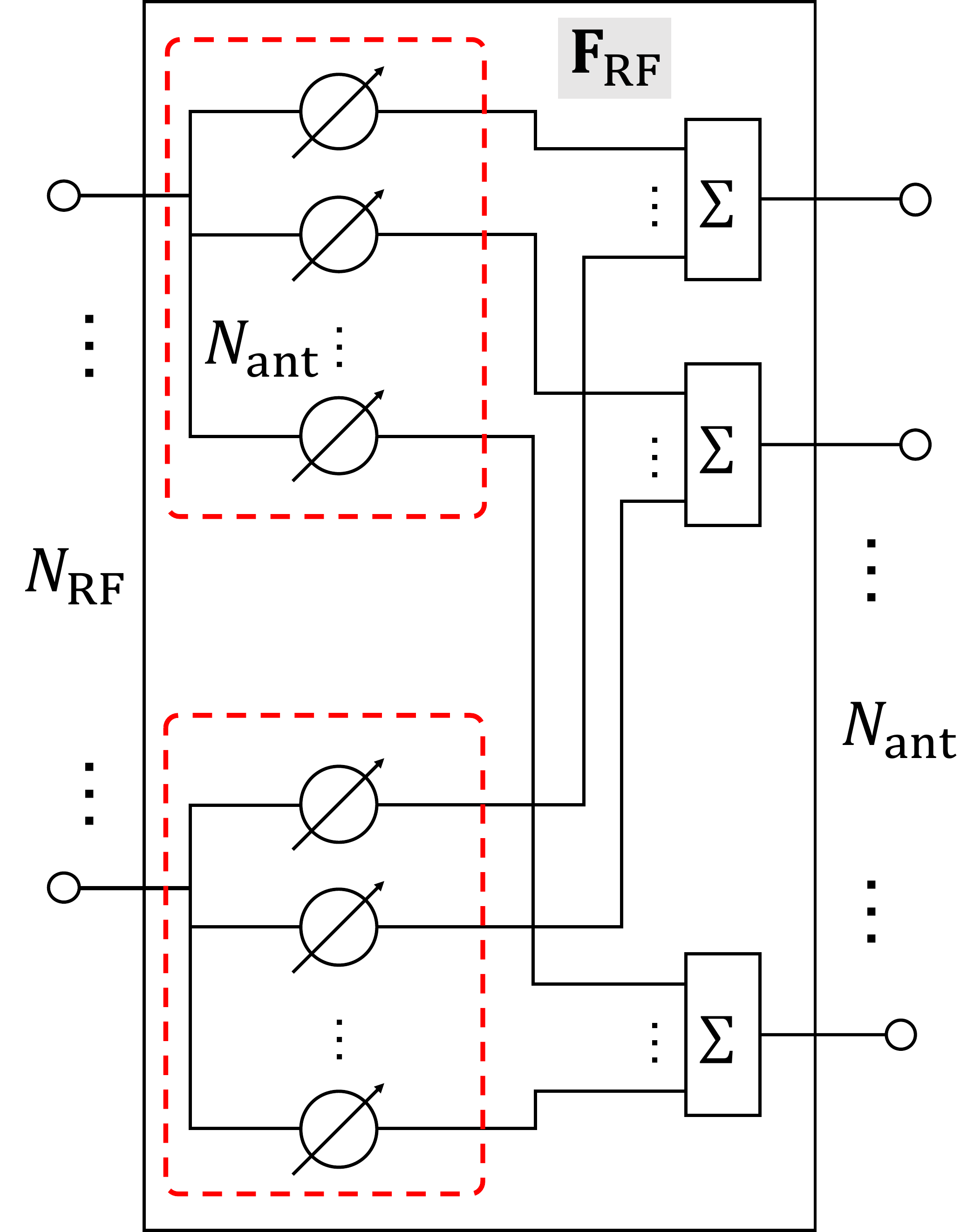}
  \caption{A diagram of the phase shifting RF stage of the precoder.}
  \label{ch4_fig:ps_precoder}
\end{figure}

In downlink, a gNB with $N_{\mathrm{ant}}^{\mathrm{BS}}$ transmit antennas and $N_{\mathrm{RF}}^{\mathrm{BS}}$ transmit RF chains and a UE with $N_{\mathrm{ant}}^{\mathrm{UE}}$ receive antennas and $N_{\mathrm{RF}}^{\mathrm{UE}}$ receive RF chains are considered. Since our focus is on HB architectures, the following inequalities hold: $N_{\mathrm{RF}}^{\mathrm{BS}}<N_{\mathrm{ant}}^{\mathrm{BS}}$ and $N_{\mathrm{RF}}^{\mathrm{UE}}<N_{\mathrm{ant}}^{\mathrm{UE}}$. Then the multipath channel matrix can be expressed as
\begin{align}
  \mathbf{H}(\tau) &= 
    \gamma \sum_{i=1}^{N_{cl}} \sum_{l=1}^{N_{ray}}
    \mathbf{H}_{i,l} \delta(\tau-\tau_{i,l})
    \in \mathbb{C}^{N_{\mathrm{ant}}^{\mathrm{UE}} \times N_{\mathrm{ant}}^{\mathrm{BS}}}
    \nonumber
    \\
  &=
    \gamma \sum_{i=1}^{N_{cl}} \sum_{l=1}^{N_{ray}}
    \alpha_{i,l} \delta(\tau-\tau_{i,l})
    \mathbf{a}_{\mathrm{UE}} (\theta_{i,l}) 
    \mathbf{a}_{\mathrm{BS}}^{\mathsf{H}}(\phi_{i,l}) 
    \nonumber
    ,
\end{align}
where $\gamma = \sqrt{N_{\mathrm{ant}}^{\mathrm{UE}} N_{\mathrm{ant}}^{\mathrm{BS}} / {N_{cl}} N_{ray} }$ is the normalization factor, 
$N_{cl}$ and $N_{ray}$ denote the number of mutipath clusters and the number of rays in each cluster, respectively, 
$\alpha_{i,l} \sim \mathcal{CN}(0,\sigma_{\alpha}^2)$ is the $(i,l)$-th path complex gain with $\sigma_{{\alpha}}^2 = 1$,
$\tau_{i,l}$ is the $(i,l)$-th path propagation delay,
and
$\theta_{i,l}$ and $\phi_{i,l}$ are the angles of arrival and departure (AoA and AoD) associated with the $(i,l)$-th path, respectively. 
$\delta(\cdot)$ denotes the impulse function, and $(\cdot)^{\mathsf{H}}$ denotes the conjugate transpose matrix.
The $(i,l)$-th path AoA and AoD are further defined as 
$\theta_{i,l} = \theta_{i} + \Delta \theta_{i,l}$ and 
$\phi_{i,l} = \phi_{i} + \Delta \phi_{i,l}$, respectively. 
$\mathbf{a}_{\mathrm{UE}} (\theta_{i,l})$ and $\mathbf{a}_{\mathrm{BS}}(\phi_{i,l})$ are the receive and transmit array response vectors at the given angles. 
Assuming uniform linear arrays (ULAs) with half-wavelength spacing for both transmit and receive antennas, the transmit array response vector is given as
\begin{align}
  \mathbf{a}_{\mathrm{BS}}(\phi) &= 
    \sqrt{\frac{1}{N_{\mathrm{ant}}^{\mathrm{BS}}}} 
    \left[1, e^{j \pi \sin(\phi)}, \ldots, e^{j \pi (N_t - 1)\sin(\phi)} \right]^{\mathsf{T}}, \nonumber
\end{align}
where $(\cdot)^{\mathsf{T}}$ denotes the matrix transpose. The receive array response is similarly defined. 
By transforming into the frequency domain, the channel matrix for the $k$-th subcarrier is given by
\begin{align}
  \label{ch4_eq:H_k_1}
  \mathbf{H}[k] = 
    \gamma \sum_{i=1}^{N_{cl}} \sum_{l=1}^{N_{ray}}
    \alpha_{i,l} e^{-\frac{j 2 \pi f_s \tau_{i,l} k}{K}} 
    \mathbf{a}_{\mathrm{UE}} (\theta_{i,l}) 
    \mathbf{a}_{\mathrm{BS}}^{\mathsf{H}}(\phi_{i,l}) 
    ,
\end{align} 
where $f_s$ is the sample rate of ADCs at the UE and $K$ is the total number of subcarriers in OFDM. 
For the sake of simplicity, it is assumed that the propagation delays of rays in a given cluster are identical, i.e., $\tau_{i} = \tau_{i,l}, \forall l$.

The transmit array response matrix $\mathbf{A}_{\mathrm{BS}}$ can be constructed by collecting the transmit array responses evaluated at $N_{cl} N_{ray}$ AoDs as
\begin{align}
  \mathbf{A}_{\mathrm{BS}} =
  \begin{bmatrix}
    \mathbf{a}_{\mathrm{BS}}(\phi_{1,1}),
    \mathbf{\mathbf{a}_{\mathrm{BS}}}(\phi_{1,2}), \allowbreak
    \ldots, \allowbreak
    \mathbf{\mathbf{a}_{\mathrm{BS}}}(\phi_{N_{cl},N_{ray}})    
  \end{bmatrix}
  ,
  \nonumber
\end{align}
and the receive array response matrix $\mathbf{A}_{\mathrm{UE}}$ can also be similarly constructed. The channel matrix in \eqref{ch4_eq:H_k_1} can be rewritten with the transmit and receive array response matrices as
\begin{align}
  \label{ch4_eq:H_k_2}
  \mathbf{H}[k] = \mathbf{A}_{\mathrm{UE}} \mathbf{H}_{d}[k] \mathbf{A}_{\mathrm{BS}}^{\mathsf{H}},
\end{align}
where 
$\mathbf{H}_{d}[k] \in \mathbb{C}^{N_{cl} N_{ray} \times N_{cl} N_{ray}}$
is the diagonal matrix with the elements being the scaled complex multipath component (MPC) path gain associated with a pair of an AoD and an AoA. 

Considering a single UE, the received signal in the $m$-th block and the $k$-th subcarrier is denoted by 
$\mathbf{y}_{m}[k] \in \mathbb{C}^{N_{\mathrm{RF}}^{\mathrm{UE}} \times 1}$ 
and can be expressed as
\begin{align}
  \mathbf{y}_{m}[k] &= \sqrt{\rho} \mathbf{W}_{m}[k]^{\mathsf{H}} 
    \mathbf{H}[k] \mathbf{F}_{m}[k] \mathbf{s}_{m}[k] 
    + 
    \mathbf{W}_{m}[k]^{\mathsf{H}} \mathbf{z}_{m}[k] 
    \nonumber 
    \\
  &= 
    \sqrt{\rho} \mathbf{W}_{m}[k]^{\mathsf{H}} \mathbf{H}[k] \mathbf{x}_{m}[k] 
    + \mathbf{n}_{m}[k] 
    \nonumber
  ,
\end{align}
where
$\rho$ is the transmit power of each subcarrier, and
$\mathbf{W}_{m}[k] \in \mathbb{C}^{ N_{\mathrm{ant}}^{\mathrm{UE}} \times N_{\mathrm{RF}}^{\mathrm{UE}}}$ and
$\mathbf{F}_{m}[k] \in \mathbb{C}^{ N_{\mathrm{ant}}^{\mathrm{BS}} \times N_{\mathrm{RF}}^{\mathrm{BS}}}$ 
are the receive combiner at the UE and and the transmit precoder at the gNB, respectively. 
The combiner/precoder are a product of the RF and baseband (BB) combiner/precoder, in other words, 
$\mathbf{W}_{m}[k] = \mathbf{W}_{\mathrm{RF},m} \mathbf{W}_{\mathrm{BB}, _{m}}[k]$ and
$\mathbf{F}_{m}[k] = \mathbf{F}_{\mathrm{RF}, m} \mathbf{F}_{\mathrm{BB}, m}[k]$. Note that the RF combiner and precoder are frequency-flat whereas BB ones are frequency dependent. 
$\mathbf{s}_{m}[k] \in \mathbb{C}^{N_{\mathrm{RF}}^{\mathrm{BS}}}$ is the transmit symbol vector, and
$\mathbf{x}_{m}[k]$ denotes $\mathbf{F}_{m}[k] \mathbf{s}_{m}[k]$.
$\mathbf{z}_{m}[k] \in \mathbb{C}^{N_{\mathrm{ant}}^{\mathrm{UE}} } \sim \mathcal{CN}(0, \sigma_n^2 \mathbf{I})$ is the noise vector, and $\mathbf{n}_{m}[k]$ denotes $\mathbf{W}_{m}[k]^{\mathsf{H}} \mathbf{z}_{m}[k]$.
$k$, $m$ and $n$ denote the index of the subcarrier, the SRS, the user, respectively.
The transmit SNR is defined by $\rho / \sigma_n^2$.
Fig.~\ref{ch4_fig:system_diagram} shows a block diagram of a general hybrid beamforming architectures by visualizing the signal flows in an OFDM communication system. 

In this paper, we consider that the RF stage of the precoder and the combiner is implemented with phase shifters and that the phase shifters have a $b_{PS}$-bit resolution. Therefore, column vectors of the RF precoder $\mathbf{F}_{\mathrm{RF},m}$ are selected from a set
$\mathcal{F} = \{ \mathbf{f} \in \mathbb{C}^{N_{\mathrm{ant}}^{\mathrm{BS}}} : 
|\mathbf{f}_{i}| = \sqrt{1/N_{\mathrm{ant}}^{\mathrm{BS}}} \}, 
\angle{\mathbf{f}_{i}} \in \Theta \}$
where
$\Theta = \{ \theta : 2\pi n / 2^{b_{PS}}, n=0, \ldots , 2^{b_{PS}}-1 \}$
and
$\mathbf{f}_{i}$ denotes the $i$-th element of the vector $\mathbf{f}$. Fig.~\ref{ch4_fig:ps_precoder} illustrates a detailed view of phase shifting network based RF precoder.
\section{Sparse Formulation} 
\label{ch4_sec:sparse_formulation}
For initial access and and beam detection purposes, it is assumed that gNB uses a precoding codebook that contains $M_{\mathrm{BS}}$ precoding matrices to illuminate its coverage. While the gNB sweeps the codebook, the UE keeps one combining matrix and then switch to the next one in the subsequent gNB codebook sweep. Assuming the combining codebook has $M_{\mathrm{UE}}$ combining matrices, the UE can collect up to $M(=M_{\mathrm{BS}} M_{\mathrm{UE}})$ blocks. For simplicity, we suppose that one block contains a single OFDM symbol.
Assuming that the channel remains constant over $M$ blocks, the collected received signal vectors $\mathbf{y}_{1, \ldots, M}[k]$ at the $k$-th subcarrier can be combined into a matrix 
$\mathbf{Y}[k] \in \mathbb{C}^{ N_{\mathrm{RF}}^{\mathrm{UE}} M_{\mathrm{UE}} \times M_{\mathrm{BS}}}$ 
which can be expressed as 
\begin{align}
  \label{ch4_eq:y_m_k}
  \mathbf{Y}[k] &=
    \left[ \mathbf{\bar{y}}_1[k], \ldots, \mathbf{\bar{y}}_{M_{\mathrm{BS}}}[k] \right]
    \nonumber
    \\
  &= 
    \sqrt{\rho} \mathbf{W}[k]^{\mathsf{H}} \mathbf{H}[k] \mathbf{X}[k] + \mathbf{W}[k]^{\mathsf{H}} \mathbf{Z}[k] 
    \nonumber 
    \\
  &= 
  \sqrt{\rho} \mathbf{W}[k]^{\mathsf{H}} \mathbf{H}[k] \mathbf{X}[k] + \mathbf{N}[k]
  ,
\end{align}
where 
$\mathbf{\bar{y}}_i[k] = [ 
\mathbf{y}_{(i-1) M_{\mathrm{UE}} + 1}[k]^{\mathsf{T}}, 
\ldots, 
\mathbf{y}_{i M_{\mathrm{UE}}}[k]^{\mathsf{T}} ]^\mathsf{T}$, 
$\mathbf{W}[k] = [ \mathbf{W}_{1}[k], \mathbf{W}_{2}[k], \ldots, \mathbf{W}_{M_{\mathrm{UE}}}[k] ] 
\in \mathbb{C}^{ N_{\mathrm{ant}}^{\mathrm{UE}} \times N_{\mathrm{RF}}^{\mathrm{UE}} M_{\mathrm{UE}} }$, 
$\mathbf{X}[k] = [ \mathbf{x}_{1}[k], \mathbf{x}_{2}[k], \ldots, \mathbf{x}_{M_{\mathrm{BS}}}[k] ] 
\in \mathbb{C}^{N_{\mathrm{ant}}^{\mathrm{BS}} \times M_{\mathrm{BS}}}$,	
and 
\begin{align}
  \mathbf{N}[k] &= 
    \mathrm{blkdiag}\{\mathbf{W}_{1}[k], \ldots, \mathbf{W}_{M_{\mathrm{BS}}}[k] \}^{\mathsf{H}} \times
    \nonumber
    \\
  &
    \begin{bmatrix}
      \mathbf{n}_{1}[k] & \mathbf{n}_{M_{\mathrm{UE}}+1}[k] & \ldots & \mathbf{n}_{(M_{\mathrm{BS}}-1) M_{\mathrm{UE}}+1}[k]
      \\
      \mathbf{n}_{2}[k] & \mathbf{n}_{M_{\mathrm{UE}}+2}[k] & \ldots & \mathbf{n}_{(M_{\mathrm{BS}}-1) M_{\mathrm{UE}}+2}[k]
      \\
      \vdots & \vdots & \ddots & \vdots
      \\
      \mathbf{n}_{M_{\mathrm{UE}}}[k] & \mathbf{n}_{2 M_{\mathrm{UE}}}[k] & \ldots & \mathbf{n}_{M_{\mathrm{BS}} M_{\mathrm{UE}}}
    \end{bmatrix}
    \nonumber
    \\
  &=
    \begin{bmatrix}
      \mathbf{z}_{1}[k] & \mathbf{z}_{M_{\mathrm{BS}}+1}[k] & \ldots & \mathbf{z}_{M - M_{\mathrm{BS}} + 1}[k] \\
      \vdots & \vdots & \ddots & \vdots \\
      \mathbf{z}_{M_{\mathrm{BS}}}[k] & \mathbf{z}_{2M_{\mathrm{BS}}}[k] & \ldots & \mathbf{z}_{M}[k]
    \end{bmatrix}
    \nonumber
    .
\end{align}
By vectorizing 
both sides of \eqref{ch4_eq:y_m_k}, 
the following equation can be obtained:
\begin{align}
  \label{ch4_eq:y_k}
  \mathbf{y}[k] 
  &= 
    \sqrt{\rho} \left( \mathbf{X}[k]^{\mathsf{T}} \otimes \mathbf{W}[k]^{\mathsf{H}} \right) \mathrm{vec}\left( \mathbf{H}[k] \right) + \mathbf{n}[k] 
    \nonumber 
    \\
  &= 
    \sqrt{\rho} \mathbf{\Phi}[k] \mathrm{vec}\left( \mathbf{H}[k] \right) + \mathbf{n}[k] \in \mathbb{C}^{N_{\mathrm{RF}}^{\mathrm{UE}} M \times 1}  
  ,
\end{align}
where 
$\mathbf{\Phi}[k] = \mathbf{X}[k]^{\mathsf{T}} \otimes \mathbf{W}[k]^{\mathsf{H}}
\in \mathbb{C}^{N_{\mathrm{RF}}^{\mathrm{UE}} M 
\times N_{\mathrm{ant}}^{\mathrm{UE}} N_{\mathrm{ant}}^{\mathrm{BS}}}$
is the sensing matrix for the $k$-th subcarrier, and 
$\mathbf{n}[k] = \mathrm{vec}(\mathbf{N}[k]) \sim \mathcal{CN}(0, \sigma_n^2 \mathbf{I}_{M_{\mathrm{UE}}} \otimes \mathbf{W}[k]^{\mathsf{H}} \mathbf{W}[k])$. 
The matrix operator $\otimes$ denotes the Kronecker product, and $\mathrm{vec}(\cdot)$ is the vectorization operation. 

We define the transmit array response grid matrix by
\begin{align}
\bar{\mathbf{A}}_{\mathrm{BS}} =
  \begin{bmatrix}
    \mathbf{a}_{\mathrm{UE}}(\vartheta_{0}),
    \mathbf{a}_{\mathrm{UE}}(\vartheta_{1}), \allowbreak
    \ldots, \allowbreak
    \mathbf{a}_{\mathrm{UE}} \allowbreak (\vartheta_{G_{\mathrm{BS}}-1})
  \end{bmatrix},
  \nonumber
\end{align}
$\in \mathbb{C}^{N_{\mathrm{ant}}^{\mathrm{BS}} \times G_{\mathrm{BS}}} $
where
$G_{\mathrm{BS}} = \allowbreak N_{\mathrm{ant}}^{\mathrm{BS}} \allowbreak K_{\mathrm{BS}}$ 
is the number of transmit angle grid bins, 
$K_{\mathrm{BS}}$ is the transmit angle grid multiplier, 
and
$\vartheta_{i} = \sin^{-1}(\frac{-i}{G_{\mathrm{BS}}})$ 
is the transmit grid angle.
The receive array response grid matrix $\bar{\mathbf{A}}_{\mathrm{UE}}$ is similarly defined. Ignoring errors that may be caused by angle grid quantization, the channel matrix in \eqref{ch4_eq:H_k_2} can be rewritten as 
\begin{align}
  \mathbf{H}[k] = \bar{\mathbf{A}}_{\mathrm{\mathrm{UE}}} \bar{\mathbf{H}}[k] \bar{\mathbf{A}}_{\mathrm{BS}}^{\mathsf{H}}
  \nonumber
  , 
\end{align}
and $\mathrm{vec}(\mathbf{H}[k])$ can be expressed as
\begin{align}
  \mathrm{vec}\left(\mathbf{H}[k] \right)
  &= 
    \left( 
      \bar{\mathbf{A}}_{\mathrm{BS}}^{*} \otimes \bar{\mathbf{A}}_{\mathrm{UE}} 
    \right) 
    \mathrm{vec} \left(\bar{\mathbf{H}}[k]\right) 
    \nonumber
  \\
  &= 
    \left( 
      \bar{\mathbf{A}}_{\mathrm{UE}}^{*} \otimes \bar{\mathbf{A}}_{\mathrm{BS}} 
    \right) 
    \mathbf{h}[k] \nonumber \\
  &= 
  \mathbf{\Psi} \mathbf{h}[k] \nonumber 
  ,
\end{align}
where 
$\mathbf{\Psi} = \bar{\mathbf{A}}_{\mathrm{BS}}^* \otimes \bar{\mathbf{A}}_{\mathrm{UE}}
\in \mathbb{C}^{N_{\mathrm{ant}}^{\mathrm{BS}} N_{\mathrm{ant}}^{\mathrm{UE}} 
\times G_{\mathrm{BS}} G_{\mathrm{UE}}}$ 
is the sparsifying dictionary matrix,
$\bar{\mathbf{H}}[k]$
is the modified MPC gain matrix associated with the dictionary matrix, 
and 
$\mathbf{h}[k] = \mathrm{vec}( \bar{\mathbf{H}}[k] )$
is a sparse vector with $N_{cl} N_{ray}$ non-zero elements. 
Then the received signal in the right hand side of \eqref{ch4_eq:y_k} can be rewritten as
\begin{align}
  \label{ch4_eq:y_k_sparse}
  \mathbf{y}[k] &=
    \sqrt{\rho} \mathbf{\Phi}[k] \mathbf{\Psi} \mathbf{h}[k] + \mathbf{n}[k]
    \nonumber 
    .
    \\
  &= \sqrt{\rho} \left(  
      \tilde{\mathbf{X}}[k] \otimes \tilde{\mathbf{W}}[k]
    \right)
    \mathbf{h}[k] + \mathbf{n}[k]
    .
\end{align}

Detection of pairs of transmit and receive beams can be achieved by finding non-zero elements in the sparse vector $\mathbf{h}[k]$, and various CS methods now can be employed for this end. The sparse formulation in \eqref{ch4_eq:y_k_sparse} is for the $k$-th subcarrier and can directly be explored to detect beams. In that case, beam detection should be performed $N_{rs}$ times and combine the detected beam pairs where $N_{rs}$ denotes the number of subcarriers that can be exploited for beam detection purposes. 

Instead, the received signal vector 
$\mathbf{y}[k]$ for $k \in \{1, \ldots, \allowbreak N_{rs}\}$ 
can be concatenated based on an assumption that the transmit and receive array responses, i.e, $\mathbf{a}_{\mathrm{UE}}(\theta)$ and $\mathbf{a}_{\mathrm{BS}}(\phi)$, are common in all subcarriers. The support in $\mathbf{h}[k]$ -- which indicates angle grid points -- is common for all subcarriers and is denoted by $\mathbf{h}$ without the subcarrier index.
This is a reasonable assumption considering (1) a relatively narrow frequency span of SSB and (2) angle estimation performed by only finding indices of a few largest support.
Denoting 
$[ \mathbf{y}[1]^{\mathsf{T}}, \mathbf{y}[2]^{\mathsf{T}}, \ldots, \mathbf{y}[K]^{\mathsf{T}} ]^{\mathsf{T}}$ 
by $\mathbf{y}$, the following equation can be obtained:
\begin{align}
  \mathbf{y} &=
    [\mathbf{y}[1]^{\mathsf{T}}, \mathbf{y}[2]^{\mathsf{T}}, \ldots, \mathbf{y}[N_{rs}]^{\mathsf{T}} ]^{\mathsf{T}}
    \nonumber \\
  &= 
  	\sqrt{\rho}
    \begin{bmatrix}
      \mathbf{\Phi}[1] \\
      \vdots \\
      \mathbf{\Phi}[N_{rs}]
    \end{bmatrix}
    \mathbf{\Psi} \mathbf{h} + 
    \begin{bmatrix}
      \mathbf{n}[1] \\
      \vdots \\
      \mathbf{n}[N_{rs}]
    \end{bmatrix}
    \nonumber \\
  & = 
    \sqrt{\rho} \mathbf{\Phi} \mathbf{\Psi} \mathbf{h} + \mathbf{n} 
    \nonumber \\
  &=
    \sqrt{\rho} \bar{\mathbf{\Phi}} \mathbf{h} + \mathbf{n}
    ,
    \nonumber
\end{align}
where the sensing matrix 
$\mathbf{\Phi} = 
[\mathbf{\Phi}[1]^{\mathsf{T}}, \ldots, \mathbf{\Phi}[N_{rs}]^{\mathsf{T}}]^{\mathsf{T}}$ 
is a stack of sensing matrices of subcarriers, and 
$\mathbf{n} = 
[\mathbf{n}[1]^{\mathsf{T}}, \ldots, \mathbf{n}[N_{rs}]^{\mathsf{T}}]^{\mathsf{T}}$ 
is the concatenated noise vectors. 
$\tilde{\mathbf{X}}[k] = \mathbf{X}[k]^{\mathsf{T}} \bar{\mathbf{A}}_{\mathrm{UE}}^{*}$ and $\tilde{\mathbf{W}}[k] = \mathbf{W}[k]^{\mathsf{H}} \bar{\mathbf{A}}_{\mathrm{BS}}$
The objective is to solve the optimization problem
\begin{align}
  \min ||\mathbf{h}||_1 \text{ such that } ||\mathbf{y} - \sqrt{\rho} \mathbf{\Phi} \mathbf{\Psi} \mathbf{h}||_2 < \epsilon
  \nonumber
\end{align}
in order to find the downlink beam pairs. 

\section{Initial Access} 
\label{ch4_sec:initial_access}
3GPP NR defines the concept of synchronization signal block (SSB) that gNB periodically transmits in a bursty manner for multiple purposes including the initial access (IA). 
One SSB is composed of the primary and secondary synchronization signals (PSS and SSS), the physical broadcast channel (PBCH) and the demodulation reference signal (DMRS). 
One SSB spans on four OFDM symbols and 240 subcarriers in time and frequency domain, respectively. The burst periodicity is configurable
, i.e., {5, 10, 20, 40, 80, 160} ms, 
and is generally set to 20 ms for IA. The number of SSB in each burst is also configurable
, i.e., {4, 8, 64} 
and determined by the frequency band. In this paper, 64 SSB in a burst is assumed as we are targeting the mmWave frequencies. For more details, refer to \cite{ts38211,ts38300}.
The IA procedure consists of four stages: beam sweeping, beam measurement, beam determination and beam reporting. 
During the beam sweeping stage, at a given time, one SSB is transmitted in a single beam toward a pre-specified direction. The subsequent SSB will be transmitted in another beam so that the gNB illuminate the cell coverage. 
During beam sweeping, multiple SSB's are transmitted in different pre-specified beams to illuminate the cell coverage. 
Single beam transmission is analogous to the analog beamforming. 
In the beam measurement stage, for a given transmitted SSB, each HB architecture UE collects received signal measurements from up to $N_{\mathrm{RF}}^{\mathrm{UE}}$ directions. In the subsequent beam determination stage, various approaches can be applied to estimate the best beam.

In this paper, we compare the CS approach against the ES in beam determination. 
For the ES, 
a receive SNR that can be obtained in the beam measurements stage is used as a metric in finding beams. 
For the CS approach, the sparse formulation derived in Sec.~\ref{ch4_sec:sparse_formulation} is used and a CS algorithm is applied, and there are two precoding and combining options for this approach: the random and the deterministic. The random codebook randomly configures the phase shifters in the RF stage whereas the deterministic codebook makes use of a predetermined set of matrices for the phase shifters. In \cite{sung2020ver}, it is shown that the DFT codebook is one of those that minimize the total coherence and can better estimate mmWave channels than the random codebook does.


\begin{figure}
  \centering
  \includegraphics[width=9.5cm]{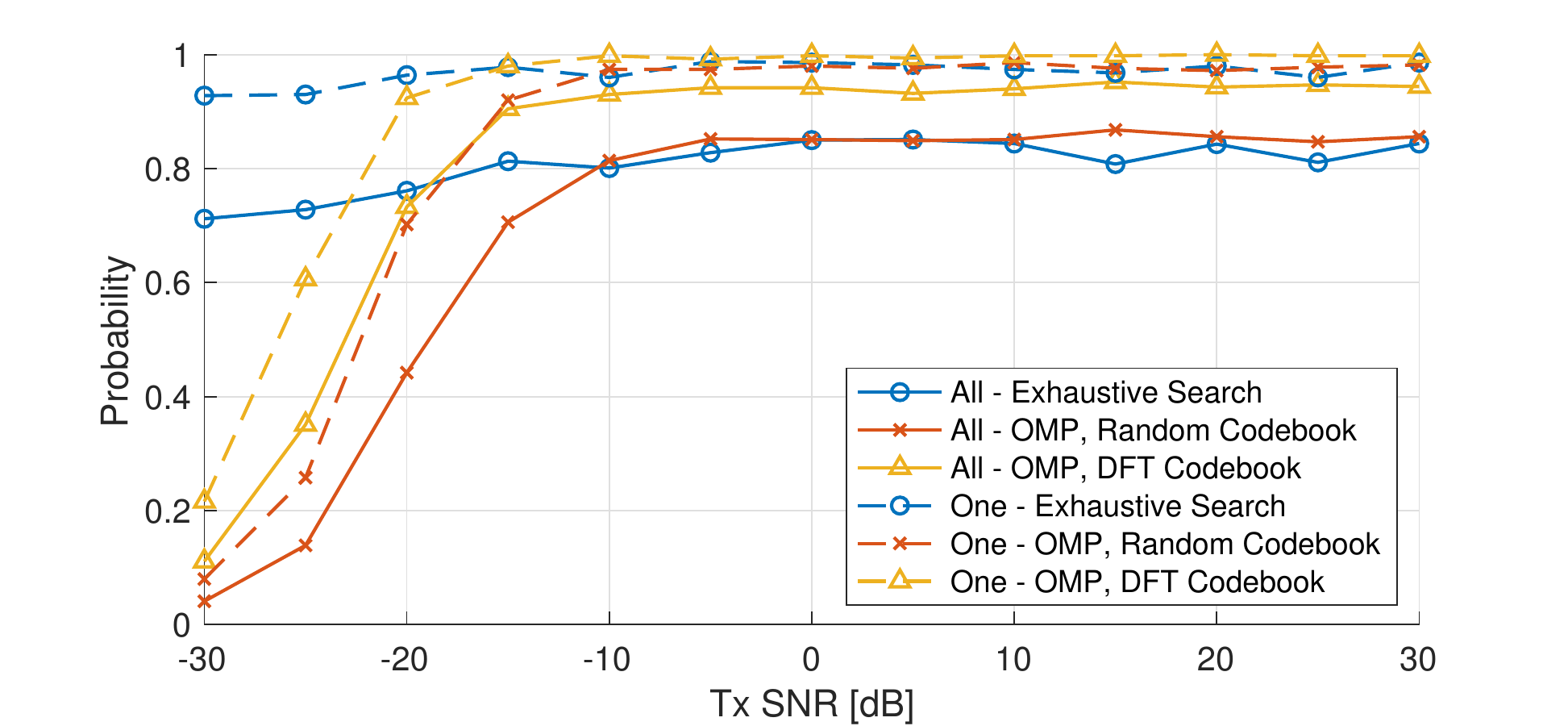}
  \caption{Beam detection probability as a function of the transmit SNR.}
  \label{ch4_fig:single_beam}
\end{figure}
\section{Numerical Results} 
\label{ch4_sec:numerical_results}
For simulation, the following parameters are used unless otherwise specified: 
$N_{\mathrm{ant}}^{\mathrm{BS}} \in \{64, 128, 256\}$, 
$N_{\mathrm{RF}}^{\mathrm{BS}}=8$,
$N_{\mathrm{ant}}^{\mathrm{UE}} = 8$, 
$N_{\mathrm{RF}}^{\mathrm{UE}}=4$,
$b_{\text{{PS}}}=6$,
$120$ kHz subcarrier spacing, 
$4096$ FFT size, 
$400$ MHz BW, 
$491.52$ MS/s sample rate,
$N_{rs}=10$ middle subcarriers,
$K_{\mathrm{BS}}=K_{\mathrm{UE}}=3$,
$M_{\mathrm{BS}}=64$,
$M_{\mathrm{UE}}=2$,
$N_{c}=2$,
and
$N_{ray}=3$.
The path delay is uniformly distributed from $0$ to $200$ ns,
the cluster means AoD and AoA are uniformly distributed in $[-\pi/2, \pi/2]$,
and
the rays have the Laplace distribution with a cluster mean and a $2^\circ$ standard deviation. For each figure, $500$ channel realizations are generated. As a representative of the CS approach, orthogonal matching pursuit (OMP) is used.
The all beam detection probability is the probability that the estimated beam pairs match the true pairs, and
the single beam detection probability is the probability that at least one estimated beam pair matches to one of the true pairs.

\begin{figure}
  \centering
  \subfloat[]{
    \includegraphics[width=4cm]{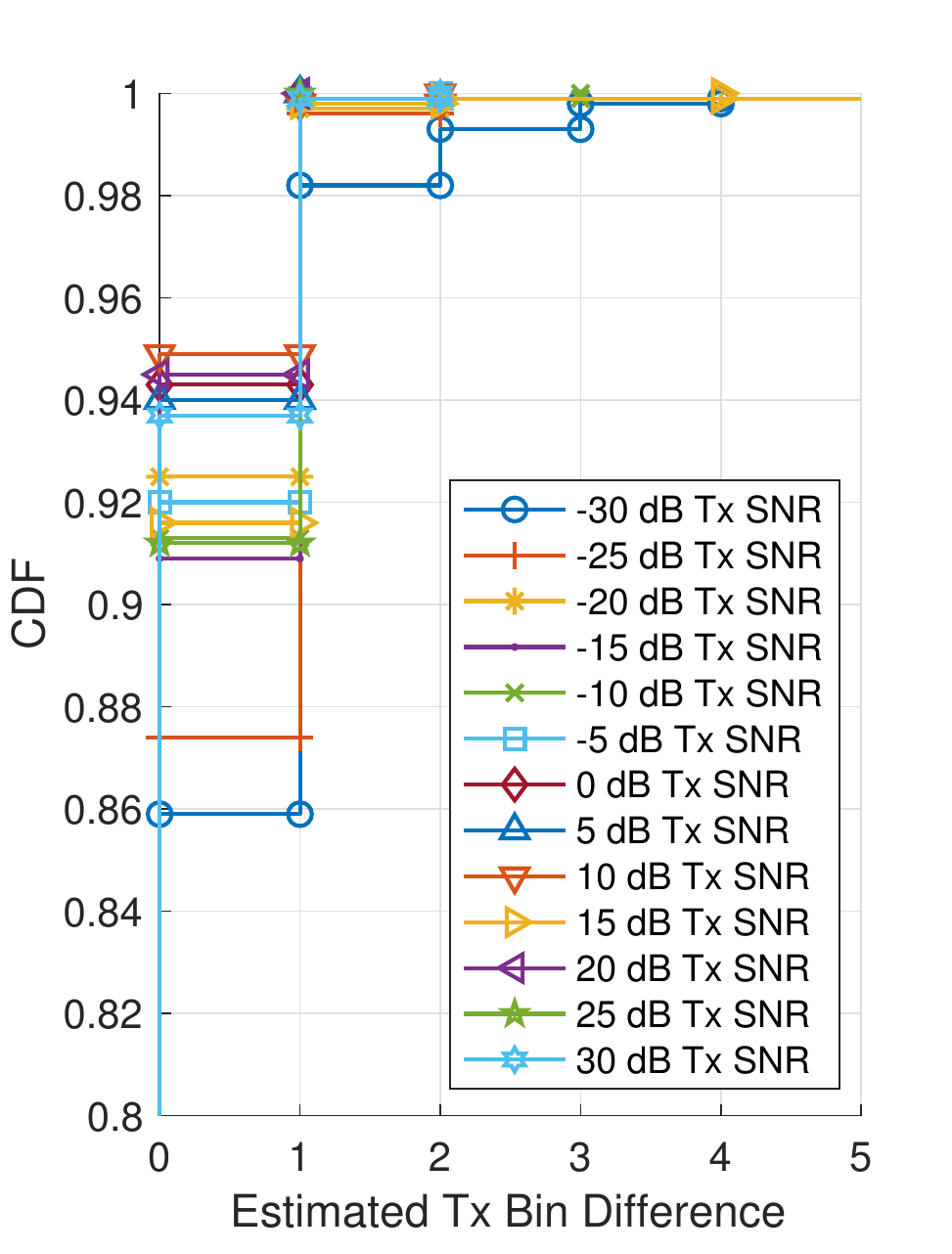}
    \label{ch4_fig:cdf_ex_tx}
  }
  \subfloat[]{
    \includegraphics[width=4cm]{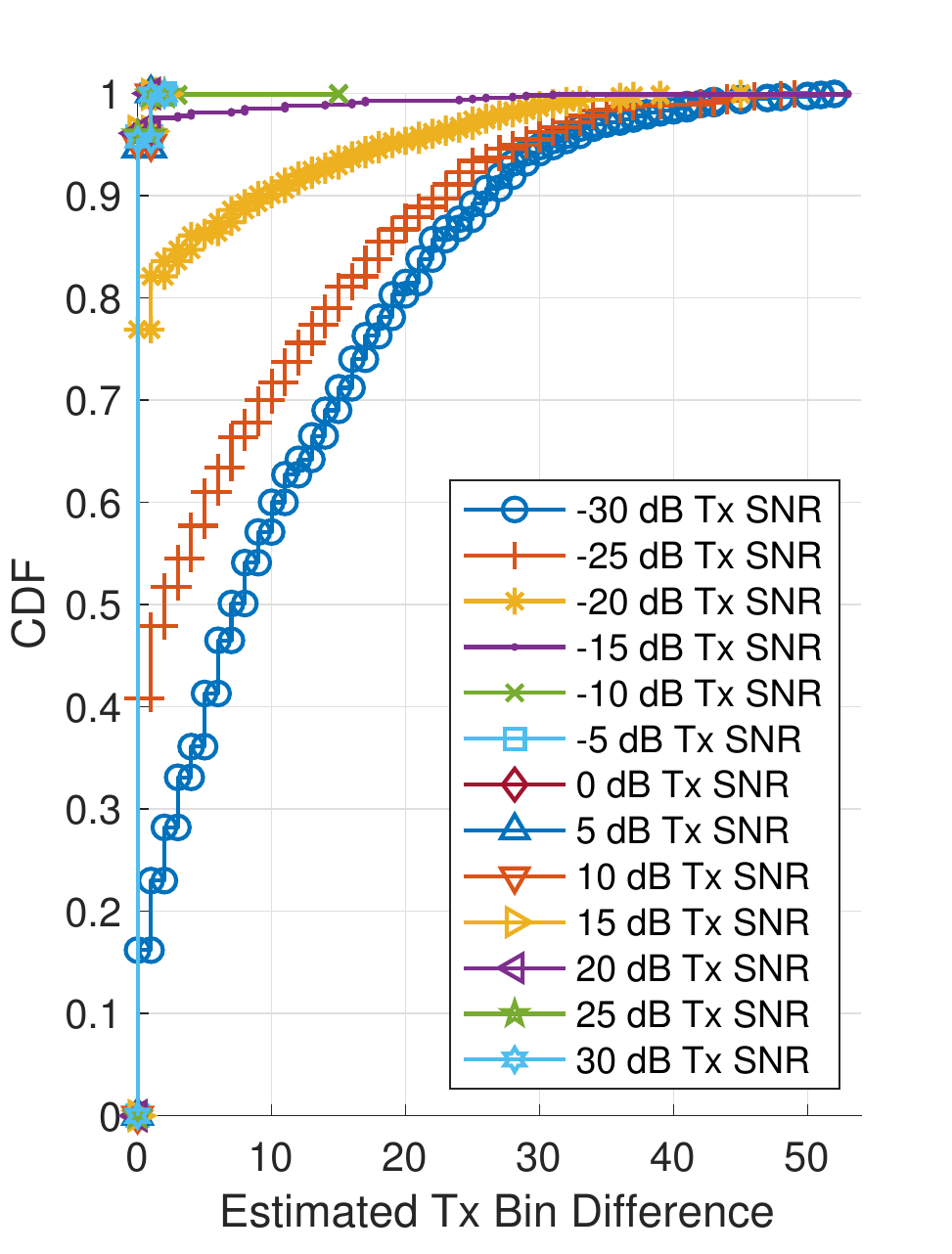}
    \label{ch4_fig:cdf_cs_tx} 
  }
  \caption{Empirical CDF of estimated transmit beam detection errors: (a) ES and (b) CS}
  \label{ch4_fig:cdf_tx}
\end{figure}
Fig.~\ref{ch4_fig:single_beam} shows both the all and single beam detection probabilities with three different approaches: the ES, OMP with the random codebook and OMP with the DFT codebook. 
For the all beam detection probability, 
OMP with the DFT codebook achieves the highest detection probability in the medium and high SNR regimes. In low SNR, the ES has higher probability than the other approaches until $-20$ dB SNR. OMP with the random codebook's probability is the lowest in the low SNR regime and is similar to that of the ES in the medium and high SNR regimes.
The figure shows that the ES is more robust to a low SNR and that high SNR is more favorable to the CS approach to better estimate the beam pairs. In addition, a smart codebook choice is critical in achieving good performance when using a CS approach. 
A similar trend is observed in the single beam detection probability with smaller performance gaps. 
All considered approaches have a very high ($>0.95$) probability with an SNR greater than $-10$ dB, and especially, OMP with the DFT codebook shows a consistently high probability. 

\begin{figure}
  \centering
  \subfloat[]{
    \includegraphics[width=4cm]{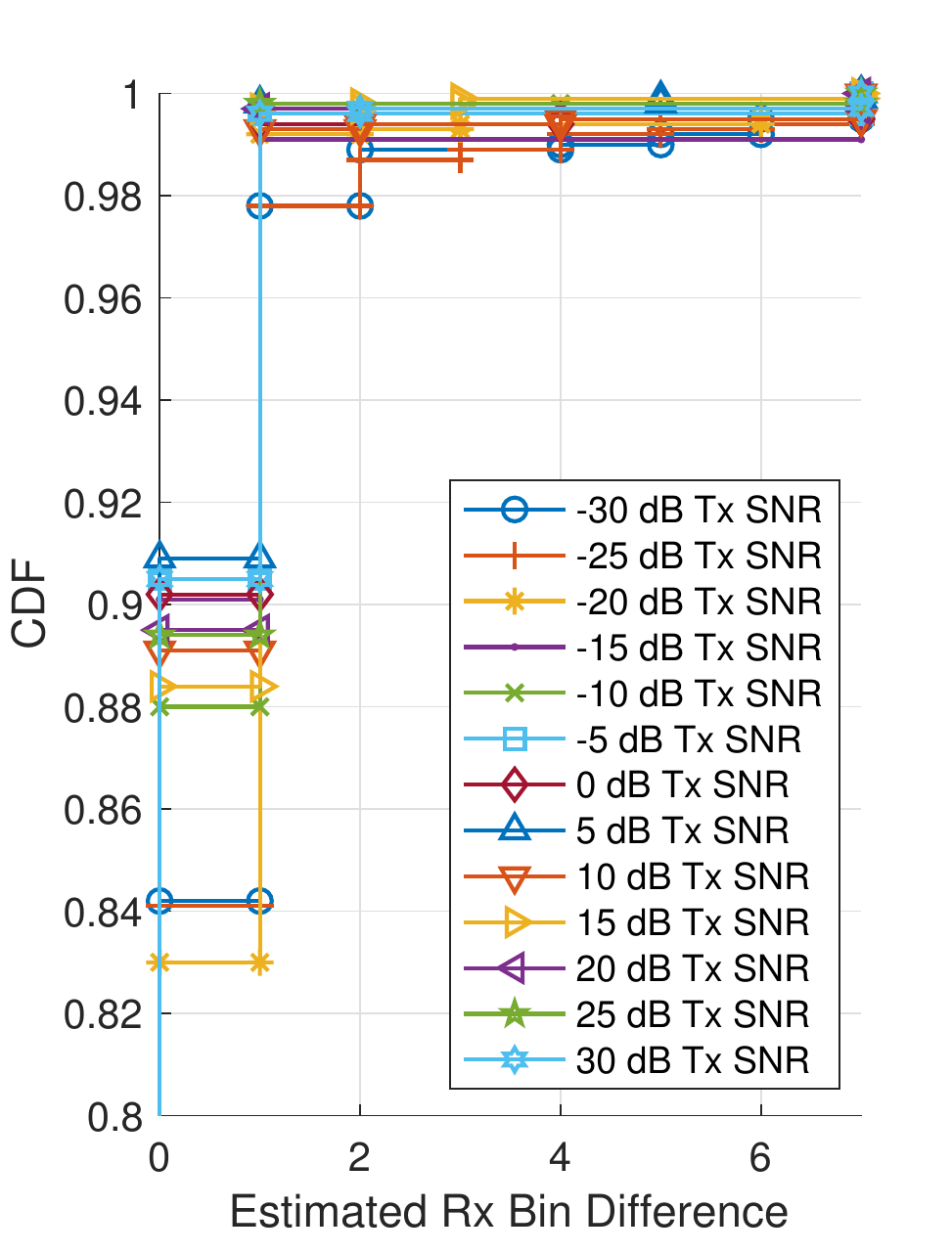}
    \label{ch4_fig:cdf_ex_rx}
  }
  \subfloat[]{
    \includegraphics[width=4cm]{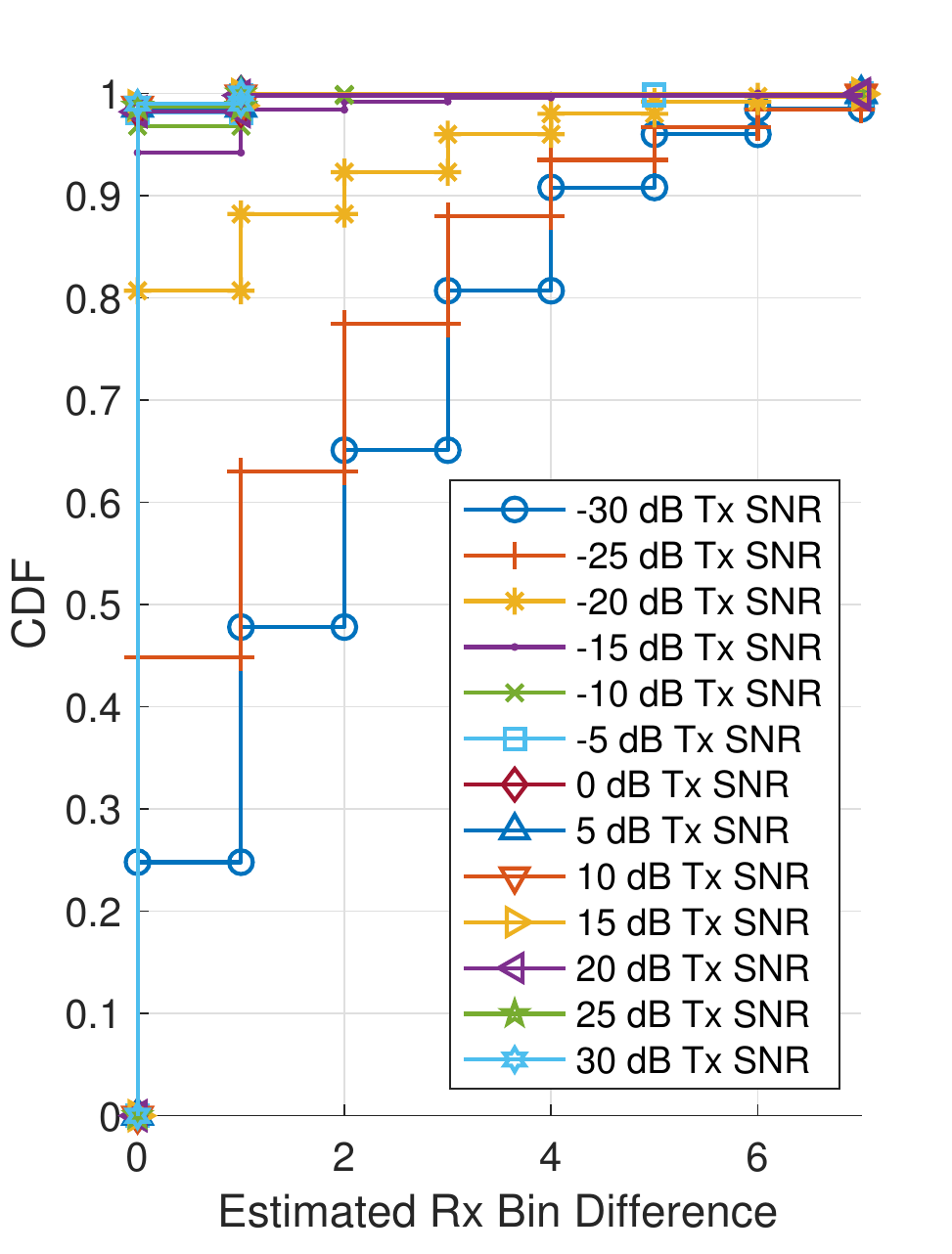}
    \label{ch4_fig:cdf_cs_rx} 
  }
  \caption{Empirical CDF of estimated receive beam detection errors: (a) ES and (b) CS}
  \label{ch4_fig:cdf_rx}
\end{figure}
We take a closer look at difference in detection probability between the ES and CS approach by plotting CDF curves in Fig.~\ref{ch4_fig:cdf_tx} and \ref{ch4_fig:cdf_rx}. They plot empirical CDF of beam detection errors in terms of beam index. 
Fig.~\ref{ch4_fig:cdf_tx} and \ref{ch4_fig:cdf_rx} are for the transmit and receive beams, respectively, and the subfigures (a) and (b) are for the ES and the CS approach, respectively. For each figure, $13$ CDF curves for a range of a transmit SNR from $-30$ to $30$ dB with an increment of $5$ dB are provided. 

The transmit beam detection errors in Fig.~\ref{ch4_fig:cdf_tx} show that the both approaches can achieve similar high detection accuracy ($\sim95\%$ at zero difference) in the high and medium SNR regimes (e.g., $-15$ to $30$ dB SNR); however, the receive beam detection accuracy from the ES is worse than the CS approach as observed in Fig.~\ref{ch4_fig:cdf_rx}. It is a primary factor that causes the performance gap between the two approaches in Fig.~\ref{ch4_fig:single_beam}. It implies that the CS-based approach is a better candidate if the system cannot tolerate any beam errors. In the low SNR regime (e.g., $-30$ to $-20$ dB SNR), the CS approach yields lower accuracy than the ES in both transmit and receive beam detection. Thus, the ES may be considered for cell-edge users. 

\begin{figure}
  \centering
  \subfloat[]{
  	\includegraphics[width=9cm]{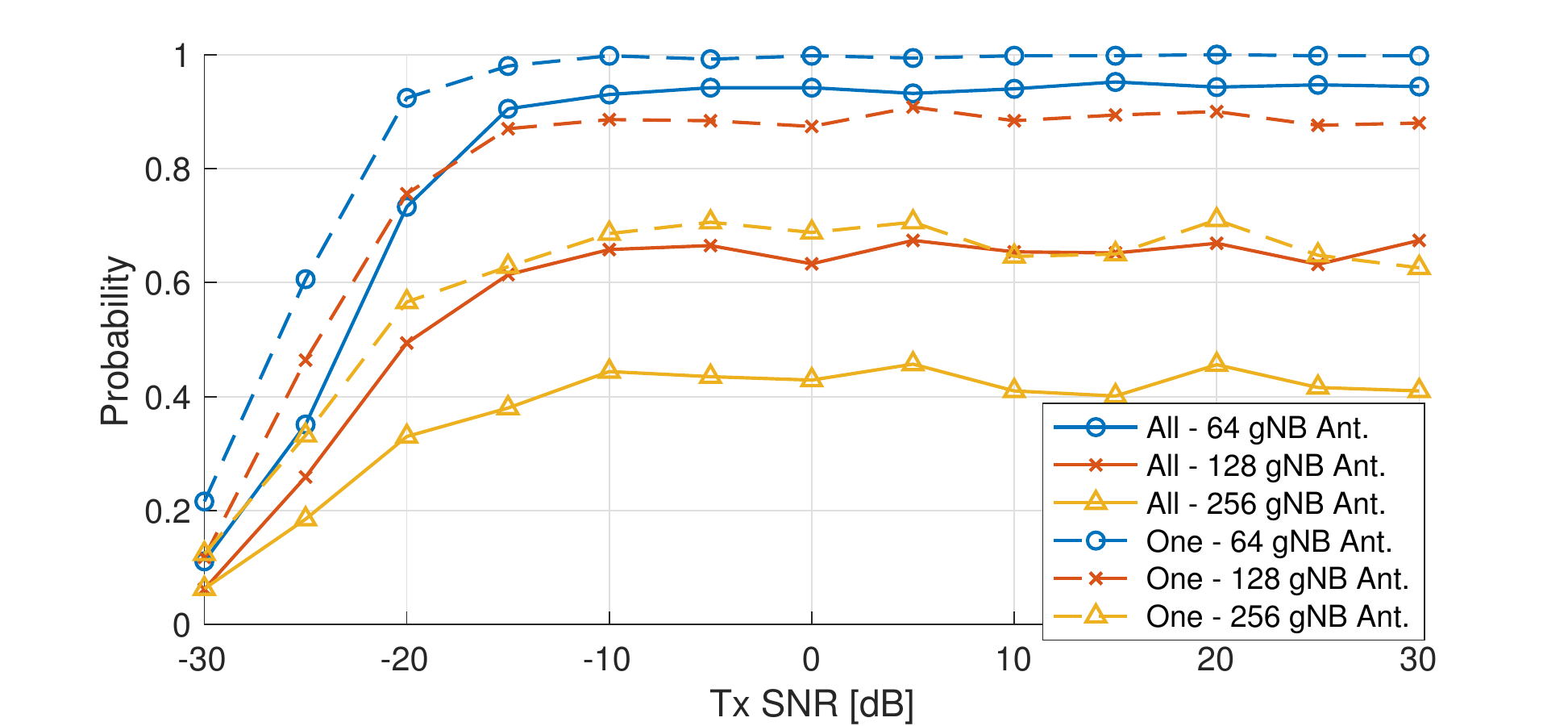}
  	\label{ch4_fig:min_beam}
  }
  \hfil
  \subfloat[]{
  	\includegraphics[width=9cm]{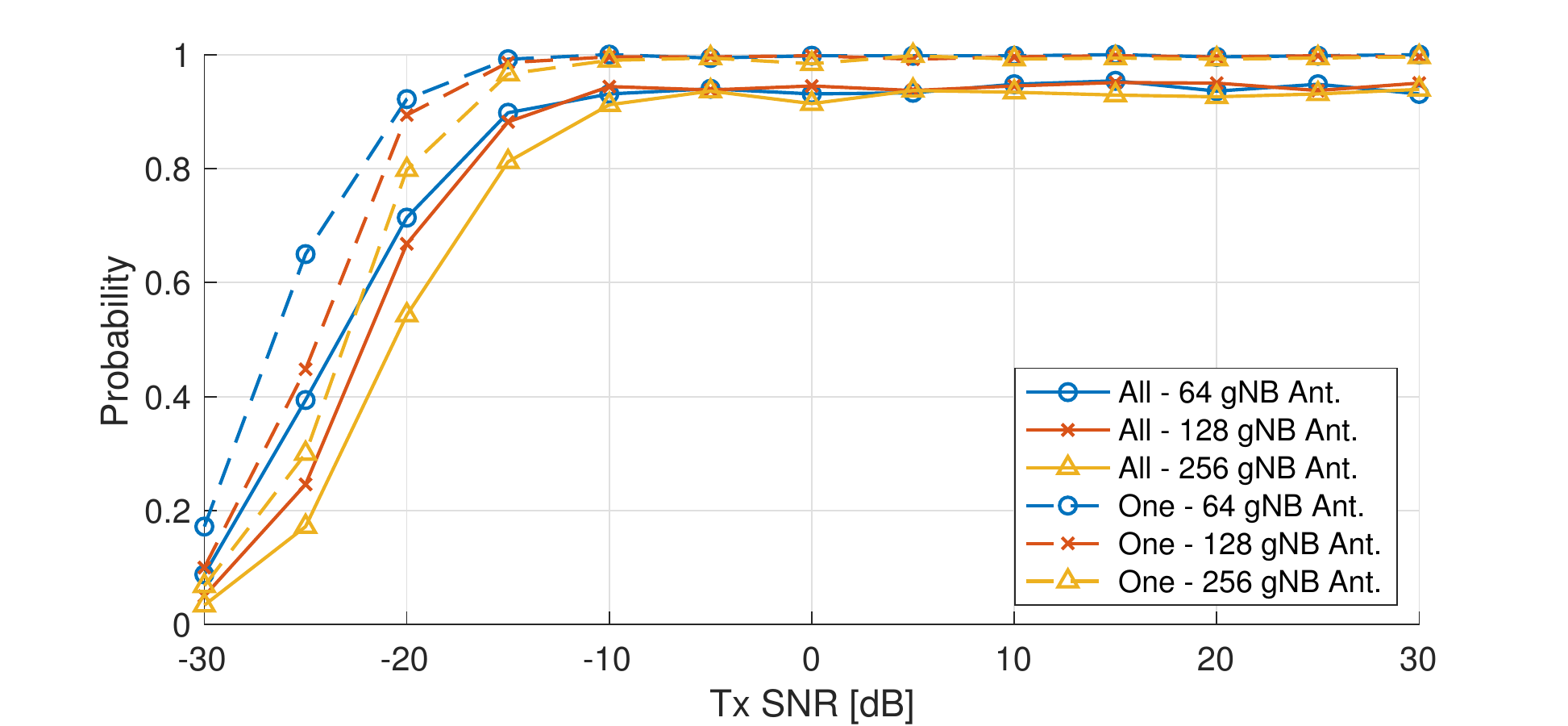}	
  	\label{ch4_fig:det_codebook}
  }
  \caption{Beam detection probability as a function of the transmit SNR: (a) DFT codebook and (b) codebook proposed in \cite{sung2020ver}} 
  \label{ch4_fig:more_antennas}
\end{figure}
Now we take more than $64$ gNB antennas into consideration: $128$ and $256$, and the beam detection probability is provided in Fig.~\ref{ch4_fig:more_antennas}. Since the maximum number of SSB in a burst is $64$ and the number of beams in the DFT codebook is the same as the transmit antennas, $128$ and $256$ antenna gNB cannot make use of the ES. 
Thus, in Fig.~\ref{ch4_fig:more_antennas}, we focus on the CS approach. 
Each SSB is transmitted using either 
(i) $N_{\mathrm{ant}}^{\mathrm{BS}}/L$ DFT beams blindly, e.g., 2 beams and 4 beams in 128 and 256 antenna systems (Fig.~\ref{ch4_fig:min_beam}) or 
(ii) all available beams using the carefully designed codebook \cite{sung2020ver} (Fig.~\ref{ch4_fig:det_codebook}). In order for further performance improvement, the deterministic ordering \cite{sung2020hyb} may be considered. 
The two codebooks achieves similar performance with $64$ antennas; however, detection probabilities are degraded with a greater number of antennas in Fig.~\ref{ch4_fig:min_beam} whereas the probabilities remain almost unchanged in Fig.~\ref{ch4_fig:det_codebook}.
Comparison of the two subfigures suggests that careful codebook design is crucial in maintaining beam detection accuracy even with a growing number of antennas. 


\section{Conclusion} 
\label{ch4_sec:conclusion}
We proposed a CS-based downlink beam detection approach for mmWave hybrid beamforming systems taking the SSB structure in the 5G NR 3GPP standard into account. 
With the exhaustive search being a benchmark, the CS approach was evaluated using the random and the DFT RF codebooks in terms of the beam detection probability. 
The simulation results showed that 
(1) the detection probability increases with transmit SNR until a saturation point, 
(2) OMP with the DFT codebook achieves the highest detection probability after saturation 
and 
(3) the exhaustive search provides the least performance degradation due to low SNR. 
We further looked into the beam detection errors to figure out the receive detection error in the exhaustive search primarily contributes the lower performance. We also considered a larger number of antennas at the gNB and explored two RF codebook options. It is observed that a smart choice of codebook is crucial in maintaining a beam detection capability with various number of antenna elements. 
\bibliographystyle{IEEEbib}
\bibliography{bib/IEEEabrv.bib,bib/paper}

\begin{thebibliography}{10}

\bibitem{ts38213}
3GPP,
\newblock ``{5G; NR; Physical layer procedures for control -- Rel. 15},''
\newblock TS 38.213, 2018.

\bibitem{giordani2019tut}
M.~{Giordani}, M.~{Polese}, A.~{Roy}, D.~{Castor}, and M.~{Zorzi},
\newblock ``A tutorial on beam management for {3GPP} {NR} at {mmWave}
  frequencies,''
\newblock {\em {IEEE} Commun. Surveys Tuts.}, vol. 21, no. 1, pp. 173--196,
  Firstquarter 2019.

\bibitem{choi2015beam}
J.~{Choi},
\newblock ``Beam selection in {mm-Wave} multiuser {MIMO} systems using
  compressive sensing,''
\newblock {\em {IEEE} Trans. Commun.}, vol. 63, no. 8, pp. 2936--2947, Aug.
  2015.

\bibitem{yan2016comp}
H.~{Yan} and D.~{Cabria},
\newblock ``Compressive sensing based initial beamforming training for massive
  {MIMO} millimeter-wave systems,''
\newblock in {\em Proc. IEEE Global Conf. on Sig. and Inform. Process.}, Dec.
  2016, pp. 620--624.

\bibitem{myers2019loc}
N.~J. {Myers} and R.~W. {Heath},
\newblock ``Localized random sampling for robust compressive beam alignment,''
\newblock in {\em Proc. IEEE Int. Conf. Acoust., Speech, Sig. Process.}, May
  2019, pp. 4644--4648.

\bibitem{desia2014init}
V.~{Desai}, L.~{Krzymien}, P.~{Sartori}, W.~{Xiao}, A.~{Soong}, and
  A.~{Alkhateeb},
\newblock ``Initial beamforming for {mmWave} communications,''
\newblock in {\em Proc. Asilomar Conf. Sign., Sys. and Comp.}, Nov. 2014, pp.
  1926--1930.

\bibitem{wei2017exha}
L.~{Wei}, Q.~{Li}, and G.~{Wu},
\newblock ``Exhaustive, iterative and hybrid initial access techniques in
  {mmWave} communications,''
\newblock in {\em Proc. IEEE Wireless Commun. and Networking Conf.}, Mar. 2017,
  pp. 1--6.

\bibitem{gior2016init}
M.~{Giordani}, M.~{Mezzavilla}, and M.~{Zorzi},
\newblock ``Initial access in {5G} {mmWave} cellular networks,''
\newblock {\em {IEEE} Commun. Mag.}, vol. 54, no. 11, pp. 40--47, Nov. 2016.

\bibitem{ts38211}
3GPP,
\newblock ``{5G; NR; Physical channels and modulation -- Rel. 15},''
\newblock TS 38.211, 2018.

\bibitem{ts38300}
3GPP,
\newblock ``{5G; NR; Overall description -- Rel. 15},''
\newblock TS 38.300, 2018.

\bibitem{sung2020ver}
J.~Sung and B.~L. Evans,
\newblock ``Versatile compressive {mmWave} hybrid beamformer codebook design
  framework,''
\newblock in {\em Proc. IEEE Int. Conf. Computing, Networking and Commun.},
  Feb. 2020.

\bibitem{sung2020hyb}
J.~Sung and B.~L. Evans,
\newblock ``Hybrid beamformer codebook design and ordering for compressive
  {mmWave} channel estimation,''
\newblock in {\em Proc. IEEE Int. Conf. Computing, Networking and Commun.},
  Feb. 2020.

\end{thebibliography}
\end{document}